\documentclass[aip,pop,amsmath,amssymb,showpacs,reprint,floatfix,lengthcheck]{revtex4-1}
\usepackage[colorlinks,linkcolor=blue,citecolor=blue,bookmarks,pdfstartview=FitH]{hyperref}
\usepackage{graphicx}
\usepackage{dcolumn}
\usepackage{bm}
\usepackage{url}
\usepackage{amssymb}
\usepackage{enumerate}
\usepackage{enumitem}

\begin{document}

\preprint{AIP/123-QED}

\title{\textcolor{blue}{Nonlinear particle simulation of parametric decay
instability of ion cyclotron wave in tokamak}}

\author{A. Kuley}
\email{akuley@uci.edu}
\affiliation{Department of Physics and Astronomy, University of California Irvine, CA 92697,USA}
\author{Z. Lin}
\affiliation{Department of Physics and Astronomy, University of California Irvine, CA 92697,USA}
\author{J. Bao}
\affiliation{Fusion Simulation Center, Peking University, Beijing 100871, China}
\affiliation{Department of Physics and Astronomy, University of California Irvine, CA 92697,USA}
\author{X. S. Wei}
\affiliation{Institute for Fusion Theory and Simulation, Zhejiang University, Hangzhou 310027, China}
\author{Y. Xiao}
\affiliation{Institute for Fusion Theory and Simulation, Zhejiang University, Hangzhou 310027, China}
\author{W. Zhang}
\affiliation{Institute of Physics, Chinese Academy of Sciences, Beijing 100190, China}
\author{G. Y. Sun}
\affiliation{Department of Physics, Institute of Theoretical Physics and Astrophysics, Xiamen University, Xiamen 361005, China}
\author{N. J. Fisch}
\affiliation{Department of Astrophysical Sciences, Princeton University, Princeton, New Jersey 08540, USA}
\affiliation{Princeton Plasma Physics Laboratory, Princeton, New Jersey 08543, USA}
\date{\today}

\begin{abstract}
Nonlinear simulation model for radio frequency (RF) waves in fusion plasmas
has been developed and verified using fully kinetic ion and drift kinetic 
electron. Ion cyclotron motion in the toroidal geometry is implemented 
using Boris push in the Boozer coordinates. Linear dispersion relation and
nonlinear particle trapping are verified for the lower hybrid (LH) wave and
ion Bernstein wave (IBW). Parametric decay instability is observed where a
large amplitude pump wave decays into an IBW sideband and an ion cyclotron
quasimode (ICQM). The ICQM induces an ion perpendicular heating with a
heating rate proportional to the pump wave intensity. 

\end{abstract}

\maketitle

\section{Introduction}
Magnetic fusion devices rely on the radio frequency (RF) waves for driving 
current and heating the plasmas, ever since it was predicted that the
power dissipated by high phase velocity waves could be 
much smaller than previously thought.\cite{PhysRevLett.41.873}
Now there are many methods of current drive considered in present-day 
tokamaks \cite{RevModPhys.59.175,0029-5515-47-6-S06,0029-5515-39-12-306}
and for the future burning plasma experiment ITER.\cite{iter} The linear theory of
RF waves using the eigen value solvers like
AORSA\cite{:/content/aip/journal/pop/8/5/10.1063/1.1359516} and 
TORIC\cite{0741-3335-41-1-002} are widely used to explain the physical phenomena in experiments. 
In spite of this, there are important situations when
linear physics fails and nonlinear phenomena like ponderomotive effects, parametric decay instability (PDI) 
can become important. The presence of PDIs have observed in several fusion devices, including
DIII-D,\cite{0029-5515-33-5-I08}
 Alcator C-Mod,\cite{:/content/aip/journal/pop/9/4/10.1063/1.1456531} 
 HT-7,\cite{0741-3335-43-9-305} 
 NSTX,\cite{:/content/aip/journal/pop/12/5/10.1063/1.1871953} FTU,\cite{Cesario10}
 ASDEX,\cite{0029-5515-32-2-I10} JT-60,\cite{Fujii1990139} and 
 EAST.\cite{:/content/aip/journal/pop/21/6/10.1063/1.4883640}  
Nonlinear phenomena of the RF waves have been studied 
theoretically\cite{:/content/aip/journal/pop/16/3/10.1063/1.3080744,
:/content/aip/journal/pop/17/6/10.1063/1.3442745,
:/content/aip/journal/pop/17/7/10.1063/1.3454692,Liu86} and numerically in the slab or cylinder
geometries with particle codes such as 
GeFi,\cite{:/content/aip/journal/pop/20/6/10.1063/1.4812196}
Vorpal,\cite{:/content/aip/journal/pop/20/1/10.1063/1.4776704,0029-5515-55-6-063002}
G-gauge.\cite{:/content/aip/journal/pop/16/3/10.1063/1.3097266}

Thus, given the importance of RF for steady state operation, instability control,
and requisite central heating we are developing a global nonlinear toroidal particle
simulation model to study the nonlinear physics associated with RF 
heating and current drive using the gyrokinetic toroidal code (GTC).\cite{Lin98}
GTC has been verified for electrostatic RF waves,\cite{:/content/aip/journal/pop/20/10/10.1063/1.4826507,
0741-3335-56-9-095020} energetic particle driven
Alfven eigenmodes,
\cite{PhysRevLett.101.095001,PhysRevLett.109.025001,PhysRevLett.111.145003}
microturbulence,\cite{PhysRevLett.103.085004,:/content/aip/journal/pop/19/3/10.1063/1.3686148}
microscopic MHD modes driven by pressure gradients and equilibrium 
currents.\cite{:/content/aip/journal/pop/21/12/10.1063/1.4905073,
:/content/aip/journal/pop/21/12/10.1063/1.4905074}
The principle advantage of the initial value approach in GTC simulation is 
that it retains all the nonlinearities and other 
physical properties (all harmonics, finite Larmor radius effects, etc.) of the RF waves. 
As a first step in developing this nonlinear toroidal particle simulation model, in the
present paper we have extended our fully kinetic ion simulation model 
from cylindrical geometry to 
the toroidal geometry.\cite{:/content/aip/journal/pop/20/10/10.1063/1.4826507,Wei2015} 
Most recently GTC has verified for the linear and nonlinear electromagnetic
simulation of lower hybrid (LH) wave.
The LH wave propagation, mode conversion, and absorption have been
simulated using fluid ion and drift kinetic electron in toroidal geometry.\cite{Jian2015a,Jian2015b}
Further developments of 
this electromagnetic model will enable us to analyze the nonlinear physics in the plasma
edge as well as its propagation to the core region.

In this paper we have implemented ion cyclotron motion in magnetic coordinates 
and verified  
linear physics  and nonlinear particle trapping for electrostatic LH wave, and ion Bernstein wave (IBW) 
using fully kinetic ion and drift kinetic electron.
We also carried out the three wave coupling in the ion cylotron
heating regime, in which the pump wave decays into an IBW sideband and an
ion cyclotron quasimode. When the frequency matching condition is satisfied, the quasimode is
strongly damped on the ion, and the ion heating takes place only in the perpendicular direction.
This quasimode induced ion heating rate is proportional to the intensity of the pump wave.

The paper is organized as follows: the physics model of fully kinetic ion and drift kinetic
electron in toroidal geometry is described in Sec.II. Section III gives the linear 
verification of the GTC simulation of the electrostatic normal modes in uniform plasmas and
nonlinear wave trapping of electron.  
Sec. IV. describes the nonlinear ion heating due to PDI. Section V summarizes this work.

\section{Physics Model in toroidal geometry}
The physics model for the fully kinetic ion, drift kinetic electron dynamics
and the numerical methods associated with the time advancement of the physical quantities
(ion position, electron guiding center, particle weight, electric field, etc.)
are described in the following sections.
\subsection{Physics Model}
\textbf{Coordinate system -}
In GTC we use toroidal magnetic coordinates $(\psi,\theta,\zeta)$ to represent the electromagnetic fields
and the plasma profile in the closed flux surface, where $\psi$ is the poloidal flux function, $\theta$ and $\zeta$ are the poloidal and
toroidal angle, respectively. The contravariant representation of the magnetic field 
is\cite{:/content/aip/journal/pof1/27/10/10.1063/1.864527}

\begin{equation}
 \vec{B}=g\nabla\zeta+I\nabla\theta,
\end{equation}
covariant representation is
\begin{equation}
 \vec{B}=q\nabla\psi\times\nabla\theta-\nabla\psi\times\nabla\zeta,
\end{equation}
and the Jacobian of this magnetic toroidal system can be written as
\begin{equation}
 J^{-1}=\nabla\psi\cdot\nabla\theta\times\nabla\zeta=\frac{B^2}{gq+I}
\end{equation}
Although GTC is capable of general toroidal geometry,\cite{:/content/aip/journal/pop/22/2/10.1063/1.4908275}
we consider a concentric cross-section tokamak in this paper. Then the radial coordinate
$\psi$ is
simplified as the minor radius $r$.
The toroidal coordinate system relates to the standard Cartesian system as follows [cf. Fig.1]
\begin{eqnarray}
 x=(R_0+r\text{cos}{\theta})\text{cos}\zeta,\nonumber\\
 y=-(R_0+r\text{cos}{\theta})\text{sin}\zeta,\\
 z= r\text{sin}\theta.\nonumber
\end{eqnarray}

By defining a covariant basis $\vec{e}_\psi=\partial_\psi\vec{r}$,
$\vec{e}_\theta=\partial_\theta\vec{r}$, $\vec{e}_\zeta=\partial_\zeta\vec{r}$,
and contravariant basis $\vec{e}^\psi=\nabla\psi$, $\vec{e}^\theta=\nabla\theta$, 
$\vec{e}^\zeta=\nabla\zeta$ the 
velocity and the electric field can be written as
\begin{equation}
 \vec{v}=v^{\psi}\vec{e}_\psi+v^{\theta}\vec{e}_\theta+v^{\zeta}\vec{e}_\zeta,
\end{equation}

\begin{figure}
\centering
\includegraphics[width=0.5\textwidth ]{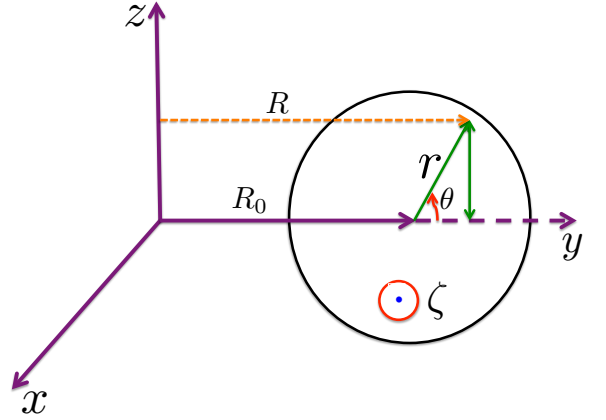}
\caption{\label{fig:epsart} Schematic diagram of the coordinates of a
concentric cross section tokamak.}
\end{figure}

\begin{equation}
 \vec{E}=-\nabla\phi=-\biggl[\frac{\partial\phi}{\partial\psi}\nabla\psi+\frac{\partial\phi}{\partial\theta}\nabla\theta+
 \frac{\partial\phi}{\partial\zeta}\nabla\zeta\biggr],
\end{equation}

where
\begin{eqnarray}
 v^{\psi}=\dot{\psi},\quad v^{\theta}=\dot{\theta}, \quad v^{\zeta}=\dot{\zeta}r\text{cos}\theta/(R_0+r\text{cos}\theta),\nonumber\\
 \vec{e}_\psi=[\text{cos}\theta\text{cos}\zeta\hat{x}-\text{cos}\theta\text{sin}\zeta\hat{y}+
 \text{sin}\theta\hat{z}](\partial r/\partial\psi),\nonumber\\
 \vec{e}_\theta=-r\text{sin}\theta\text{cos}\zeta\hat{x}+r\text{sin}\theta\text{sin}\zeta\hat{y}+r\text{cos}\theta\hat{z},\\
 \vec{e}_\zeta=-(R_0+r\text{cos}\theta)\text{sin}\zeta\hat{x}-(R_0+r\text{cos}\theta)\text{cos}\zeta\hat{y}\nonumber\\
 \psi(r)=\int_0^r(r/q)dr\nonumber\quad\quad\quad\quad\quad\quad\quad\quad
\end{eqnarray}

\textbf{Ion dynamics -}
Ion dynamics is described by the six dimensional Vlasov equation,
 \begin{equation}
\biggl[\frac{\partial}{\partial t}+\vec{v}\cdot \nabla+\frac{Z_i}{m_i}(\vec{E}+
\vec{v}\times \vec{B})\cdot\frac{\partial}{\partial \vec{v}}\biggr]f_i=0,
\end{equation}
where $f_i$ is the ion distribution function, $Z_i$ is the ion charge, and $m_i$ is the ion mass.

The evolution of the ion distribution function $f_i$ can be described by the Newtonian equation of motion in 
the presence of self-consistent electromagnetic field as follows
\begin{equation}
 \frac{d}{dt}\vec{r}=\vec{v},\quad\quad
 \frac{d}{dt}\vec{v}=\frac{Z_i}{m_i}\biggl[\vec{E}+\vec{v}\times\vec{B}\biggr]
\end{equation}
In our simulation we compute the marker particle trajectory [Eq.(9)] by the time centered Boris push
method{\cite{Boris,Birdsall,:/content/aip/journal/pop/20/10/10.1063/1.4826507} as discussed 
in the following section. 

In our GTC simulation we have implemented both 
perturbative $(\delta f)$ and non-perturbative (full-$f$) methods.
We use $(\delta f_i)$ method to reduce the particle noise. 
Now we decompose the distribution function $(f_i)$ into its
equilibrium $(f_{0i})$ and perturb part $(\delta f_i)$ i.e., $f_i=f_{0i}+\delta f_i$.
The perturbed density for ion is defined as the
fluid moment of ion distribution function, $\delta n_i=\int \delta f_i d^3v$.
By defining the particle weight $w_i=\delta f_i/f_i$, we
can rewrite the Vlasov equation for Maxwellian
ion with uniform temperature $T_i$ and uniform density as follows
\begin{equation}
 \frac{d}{dt}w_i=-\frac{Z_i}{T_i}(1-w_i)\biggl[\frac{\partial\phi}{\partial\psi}v^\psi+
 \frac{\partial\phi}{\partial\theta}v^\theta+\frac{\partial\phi}{\partial\zeta}v^\zeta\biggr]
\end{equation}

\textbf{Electron dynamics -}
Electron dynamics is described by the five-dimensional drift kinetic equation
\begin{equation}
 \biggl[\frac{\partial}{\partial t}+\dot{\vec{X}}\cdot\nabla
 +\dot{v}_\parallel\frac{\partial}{\partial v_\parallel}\biggr]f_e(\vec{X},v_\parallel,\mu,t)=0
\end{equation}
where $f_e$ is the guiding center distribution function, $\vec{X}(\psi,\theta,\zeta)$ is the 
guiding center position, $\mu$ is the magnetic moment, and $v_\parallel$ is the parallel
velocity. The evolution of the electron distribution function can be described
by the following equations of guiding center motion:\cite{RevModPhys.79.421}

\begin{eqnarray}
 \dot{\vec{X}}=v_\parallel\hat{b}+\vec{v}_E+\vec{v}_c+\vec{v}_g,\nonumber\\
 \dot{v}_\parallel=-\frac{1}{m_e}\frac{\vec{B}^*}{B}\cdot(\mu\nabla B-e\nabla\phi),
\end{eqnarray}
where $\vec{B}^*=\vec{B}+Bv_\parallel/\omega_{ce}\nabla\times\hat{b}$, and $\mu=m_ev_\perp^2/2B$.
The $\vec{E}\times\vec{B}$ drift velocity $\vec{v}_E$, the grad-$B$ drift 
velocity $\vec{v}_g$, and curvature drift velocity $\vec{v}_c$ are given by
\begin{eqnarray}
 \vec{v}_E=\frac{c\hat{b}\times\nabla\phi}{B},\nonumber\\
 \vec{v}_g=\frac{\mu}{m\omega_{ce}}\hat{b}\times\nabla B,\\
 \vec{v}_c=\frac{v_\parallel^2}{\omega_{ce}}\nabla\times \hat{b}.\nonumber
\end{eqnarray}

This electron model is suitable for the dynamics with the wave frequency 
$\omega\ll\omega_{ce}$ and $k_{\perp}\rho_e\ll 1$, where $\omega_{ce}$ is the 
electron cyclotron frequency and $\rho_e$ is the electron gyro radius.
Electron dynamics are described by conventional Runge-Kutta method.
The perturbed density for electron also can be found from the 
fluid moment of electron distribution function, $\delta n_e=\int \delta f_e d^3v$.
The weight equation for electron can be written as
\begin{equation}
 \frac{d}{dt}w_e=(1-w_e)\biggl[-e\frac{\vec{B}^*}{B_0}\cdot\nabla\phi\frac{1}{m_e}\frac{1}{f_{0e}}
 \frac{\partial f_{0e}}{\partial v_{\parallel}}\biggr]
\end{equation}
where $w_e=\delta f_e/f_{e}$ and $f_e=f_{0e}+\delta f_e$. $f_{0e}$ and $\delta f_e$ are the equilibrium and perturbed
distribution function, respectively. Simulation related to nonuniform plasma density 
and temperature will be reported in future work. \\

\textbf{Field equation -}
This paper describes the electrostatic model of fully kinetic ion
and drift kinetic electron. Most recently Bao \textit{et al}. have formulated the electromagnetic
description of this model.\cite{Jian2015a}}  The electrostatic potential can be 
calculated from the Poisson's equation
\begin{equation}
 \nabla_\perp\cdot\biggl[\biggl(1+\frac{\omega_p^2}{\omega_c^2}\biggr)\nabla_\perp\phi\biggr]
 =-4\pi(Z_i \delta n_i-e \delta n_e)
\end{equation}
Here we consider that fact that the perpendicular wavelength is much shorter
than the parallel wavelength to suppress the
high frequency electron plasma oscillation along the magnetic field line.
Second term on the left hand side corresponds to the
electron density due to its perpendicular polarization drift. For an axisymmetric system, 
the perpendicular Laplacian can be explicitly expressed as\cite{:/content/aip/journal/pop/22/2/10.1063/1.4908275}
\begin{eqnarray}
 \nabla_\perp^2=\text{g}^{\psi\psi}\frac{\partial^2}{\partial\psi^2}+
 2\text{g}^{\psi\theta}\frac{\partial^2}{\partial\psi\partial\theta_0}+
 (\text{g}^{\theta\theta}+\text{g}^{\zeta\zeta}/q^2)\frac{\partial^2}{\partial\theta_0^2}\nonumber\\
 +\frac{1}{J}\biggl(\frac{\partial J\text{g}^{\psi\psi}}{\partial\psi}+
 \frac{\partial J\text{g}^{\psi\theta}}{\partial\theta_0}\biggr)\frac{\partial}{\partial\psi}
 +\frac{1}{J}\biggl(\frac{\partial J\text{g}^{\psi\theta}}{\partial\psi}+
 \frac{\partial J\text{g}^{\theta\theta}}{\partial\theta_0}\biggr)\frac{\partial}{\partial\theta_0}\nonumber\\
\end{eqnarray}
where $\theta_0=\theta-\zeta/q$, and $\zeta_0=\zeta$. In GTC we used the field aligned coordinates $(\psi,\theta_0,\zeta_0)$,
for reducing the number of parallel grids.
Secondly we use the B-spline representation of the magnetic field, which provides
a transformation $R=R(\psi,\theta)$, and $Z=Z(\psi,\theta)$, where $(R,Z,\zeta)$ are 
the cylindrical coordinates. We define the contravariant geometric tensor 
$\text{g}^{\xi^\alpha\xi^\beta}=\nabla{\xi^\alpha}\cdot\nabla{\xi^\beta}$,
$(\xi^1,\xi^2,\xi^3)=(\psi,\theta,\zeta)$. The covariant geometric tensor  $\text{g}_{\xi^\alpha\xi^\beta}$ can be 
expressed as
\begin{eqnarray}
 \text{g}_{\psi\psi}=\biggl(\frac{\partial R}{\partial\psi}\biggr)^2+\biggl(\frac{\partial Z}{\partial\psi}\biggr)^2,\nonumber\\
 \text{g}_{\theta\theta}=\biggl(\frac{\partial R}{\partial\theta}\biggr)^2+\biggl(\frac{\partial Z}{\partial\theta}\biggr)^2,\\
\text{g}_{\psi\theta}=\frac{\partial R}{\partial\psi}\frac{\partial R}{\partial\theta}+
\frac{\partial Z}{\partial\psi}\frac{\partial Z}{\partial\theta},\nonumber
 \end{eqnarray}
and $\text{g}_{\psi\theta}=\text{g}_{\theta\psi},\text{g}_{\zeta\zeta}=R^2=(R_0+r\text{cos}\theta)^2$
for concentric cross-section tokamak, where $R_0$ is the major radius of the tokamak.


\subsection{Boris push for ion dynamics}
The efficiency of particle simulation strongly depends on the particle pusher. Boris scheme is the most 
widely used orbit integrator in explicit particle-in-cell (PIC)
simulation of plasmas. In this paper we have extended our Boris push
scheme from cylindrical geometry to toroidal geometry.\cite{:/content/aip/journal/pop/20/10/10.1063/1.4826507}
This scheme offers second order accuracy while requiring only one 
force (or field) evaluation per
step. The interplay between the PIC cycle and the Boris scheme is
schematically represented in Fig.2. 
At the beginning of each cycle the position of the particles and their
time centered velocity $\vec{v}(t-1/2)$, weight $w_i(t)$,
as well as the grid based electromagnetic fields $\vec{E}(t),\vec{B}(t)$ are given.

\begin{figure*}
\centering
\includegraphics[width=12.5cm,height=8cm]{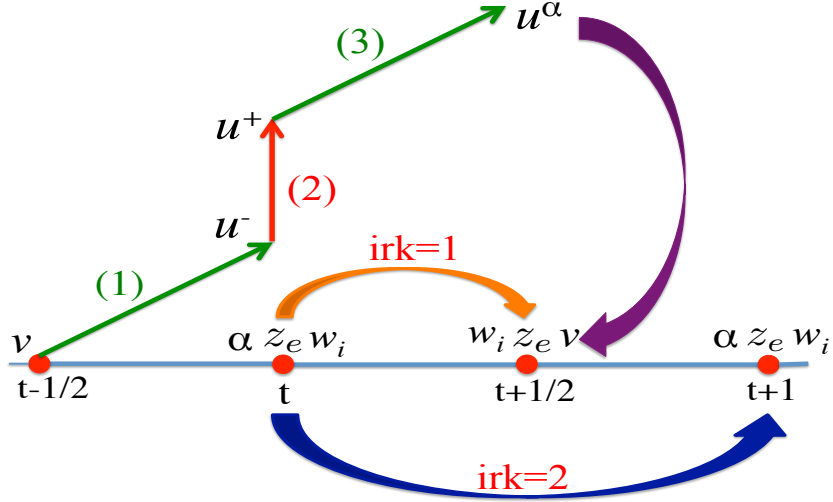}
\caption{\label{fig:epsart} Schematic diagram for the PIC cycle and ion Boris push.
The first step indicates the addition of the first half of the electric field acceleration
to the velocity $(v\rightarrow u^-)$. The second step is rotation of the velocity vector
$(u^-\rightarrow u^+)$.
In the third step,
we add the second half of the electric field impulse to the rotated
velocity component $(u^+\rightarrow u^\alpha, \alpha=\psi,\theta,\zeta)$. Ion particle weight $(w_i)$ 
and electron guiding center $(z_e)$ 
are updated using the second order Runge-Kutta (irk=1 and irk=2) method.
Dark purple blue indicates the transformation from $u^\alpha\rightarrow v$, 
which is needed to update the ion position [cf. Eq.(27)].}
\end{figure*}

In the first step, we add the first half of the electric field acceleration to the velocity $\vec{v}(t-1/2)$ to obtain the 
velocity at the particle position $\vec{r}(t)$ as follows
\begin{equation}
 \vec{u}(t)=\vec{v}(t-1/2)+\frac{Z_i}{m_i}\frac{\Delta t}{2}\vec{E}(t)
\end{equation}
One can write down the components of velocity at particle position $\vec{r}(t)$ as
\begin{eqnarray}
 u^{\alpha-}(t)=\sum_{\beta=\psi,\theta,\zeta}v^\beta(t-1/2)\vec{e}_\beta(t-1/2)\cdot\nabla\alpha(t)\nonumber\\
 +\frac{Z_i}{m_i}\frac{\Delta t}{2}\vec{E}(t)\cdot\nabla\alpha(t)
\end{eqnarray}
where $\alpha=\psi,\theta,\zeta$.  We note that in Boozer coordinates, the basis vectors are non-orthogonal in nature.
However, for simplicity we consider the orthogonal components only, and Eq.(19) can be 
rewrite as follows

\begin{eqnarray}
 u^{\psi-}(t)=\frac{1}{\text{g}_{\psi\psi}(t)}\biggl[v^\psi(t-1/2)\vec{e}_{\psi}(t-1/2)\cdot\vec{e}_{\psi}(t)\quad\quad\quad\quad\nonumber\\
+ v^\theta(t-1/2)\vec{e}_{\theta}(t-1/2)\cdot\vec{e}_{\psi}(t)\quad\quad\quad\quad\nonumber\\
 +v^\zeta(t-1/2)\vec{e}_{\zeta}(t-1/2)\cdot\vec{e}_{\psi}(t)\biggr]
 +\frac{Z_i}{m_i}\frac{\Delta t}{2}\vec{E}(t)\cdot\nabla\psi(t),\nonumber\\
 u^{\theta-}(t)=\frac{1}{\text{g}_{\theta\theta}(t)}\biggl[v^\psi(t-1/2)\vec{e}_{\psi}(t-1/2)\cdot\vec{e}_{\theta}(t)\quad\quad\quad\quad\nonumber\\
+ v^\theta(t-1/2)\vec{e}_{\theta}(t-1/2)\cdot\vec{e}_{\theta}(t)\quad\quad\quad\quad\nonumber\\
 +v^\zeta(t-1/2)\vec{e}_{\zeta}(t-1/2)\cdot\vec{e}_{\theta}(t)\biggr]
 +\frac{Z_i}{m_i}\frac{\Delta t}{2}\vec{E}(t)\cdot\nabla\theta(t),\nonumber\\
 u^{\zeta-}(t)=\frac{1}{\text{g}_{\zeta\zeta}(t)}\biggl[v^\psi(t-1/2)\vec{e}_{\psi}(t-1/2)\cdot\vec{e}_{\zeta}(t)\quad\quad\quad\quad\nonumber\\
 +v^\theta(t-1/2)\vec{e}_{\theta}(t-1/2)\cdot\vec{e}_{\zeta}(t)\quad\quad\quad\quad\nonumber\\
 +v^\zeta(t-1/2)\vec{e}_{\zeta}(t-1/2)\cdot\vec{e}_{\zeta}(t)\biggr]
 +\frac{Z_i}{m_i}\frac{\Delta t}{2}\vec{E}(t)\cdot\nabla\zeta(t)\nonumber\\
\end{eqnarray}
Using Eq.(7) we can simplify the above equations as

\begin{eqnarray}
 u^{\psi-}(t)=\frac{(\partial r/\partial\psi)_t}{\text{g}_{\psi\psi}(t)}\biggl[Av^\psi(t-1/2)(\partial r/\partial\psi)_{t-1/2}\quad\quad\nonumber\\
 +Bv^\theta(t-1/2)r(t-1/2)\quad\quad\quad\nonumber\\
+ Cv^\zeta(t-1/2)(R_0+r(t-1/2)\text{cos}\theta(t-1/2))\biggr]\nonumber\\
-\frac{Z_i}{m_i}\frac{\Delta t}{2}\frac{\partial\phi}{\partial\psi}\text{g}^{\psi\psi},\nonumber\\
u^{\theta-}(t)=\frac{1}{\text{g}_{\theta\theta}(t)}\biggl[Dv^\psi(t-1/2)(\partial r/\partial\psi)_{t-1/2}r(t)\quad\quad\nonumber\\
+Ev^\theta(t-1/2)r(t-1/2)r(t)\quad\quad\quad\nonumber\\
+Fv^\zeta(t-1/2)r(t)(R_0+r(t-1/2)\text{cos}\theta(t-1/2))\biggr]\nonumber\\
-\frac{Z_i}{m_i}\frac{\Delta t}{2}\frac{\partial\phi}{\partial\theta}\text{g}^{\theta\theta},\nonumber\\
u^{\zeta-}(t)=\frac{1}{\text{g}_{\zeta\zeta}(t)}\biggl[Gv^\psi(t-1/2)(\partial r/\partial\psi)_{t-1/2}\quad\quad\quad\quad\nonumber\\
+Hv^\theta(t-1/2)r(t-1/2)\quad\quad\quad\quad\nonumber\\
+Pv^\zeta(t-1/2)(R_0+r(t-1/2)\text{cos}\theta(t-1/2))\biggr]\nonumber\\
\times (R_0+r(t)\text{cos}\theta(t))
-\frac{Z_i}{m_i}\frac{\Delta t}{2}\frac{\partial\phi}{\partial\zeta}\text{g}^{\zeta\zeta},\quad\quad
\end{eqnarray}
where
\begin{eqnarray}
 A=P\text{cos}\theta_1\text{cos}\theta_2+\text{sin}\theta_1\text{sin}\theta_2,\nonumber\\
 B=-P\text{sin}\theta_1\text{cos}\theta_2+\text{cos}\theta_1\text{sin}\theta_2,\nonumber\\
 C=\text{cos}\theta_2(-\text{sin}\zeta_1\text{cos}\zeta_2+\text{cos}\zeta_1\text{sin}\zeta_2),\nonumber\\
 D=-P\text{cos}\theta_1\text{sin}\theta_2+\text{sin}\theta_1\text{cos}\theta_2,\nonumber\\
 E=P\text{sin}\theta_1\text{sin}\theta_2+\text{cos}\theta_1\text{cos}\theta_2,\\
 F=\text{sin}\theta_2(\text{sin}\zeta_1\text{cos}\zeta_2-\text{cos}\zeta_1\text{sin}\zeta_2),\nonumber\\
 G=\text{cos}\theta_1(-\text{cos}\zeta_1\text{sin}\zeta_2+\text{sin}\zeta_1\text{cos}\zeta_2),\nonumber\\
 P=\text{cos}\zeta_1\text{cos}\zeta_2+\text{sin}\zeta_1\text{sin}\zeta_2\nonumber\\
 H=-G\text{tan}\theta_1,\theta_1=\theta(t-1/2),\theta_2=\theta(t)\nonumber\\
 \zeta_1=\zeta(t-1/2),\zeta_2=\zeta(t)\nonumber
\end{eqnarray}

In the second step we 
consider the rotation of the velocity at time $(t)$. Rotated vector can be written as
\begin{equation}
 \vec{u}^+(t)=\vec{u}^-(t)+\vec{u}^-(t)\times\vec{s}(t)+[\vec{u}^-(t)\times\vec{T}(t)]\times\vec{s}(t)
\end{equation}
where $\vec{T}=(Z_i\vec{B}/m_i)(\Delta t/2)$ and $\vec{s}=2\vec{T}/(1+T^2)$. Most recently
Wei \textit{et al}.,\cite{Wei2015} have developed the single particle ion dynamics in general geometry using both 
toroidal and poloidal components of magnetic field. In our formulations we also incorporate
both toroidal and poloidal components of magnetic field. However, for simplicity during
rotation we consider the toroidal component of magnetic field only, 
$\vec{B}=g\nabla\zeta=q\nabla\psi\times\nabla\theta$,
and the components of the rotated vector become

\begin{eqnarray}
 u^{\psi+}(t)=\biggl[1-\biggl(\frac{2}{1+T^2}\biggr)\biggr(\frac{Z_i}{m_i}\frac{\Delta t}{2}\biggr)^2B^2\biggr]u^{\psi-}(t)\nonumber\\
+ \biggl(\frac{2}{1+T^2}\biggr)\biggr(\frac{Z_i}{m_i}\frac{\Delta t}{2}\biggr)\biggl(\frac{g}{J}\biggr)\text{g}_{\theta\theta}(t)u^{\theta-}(t),\nonumber\\
u^{\theta+}(t)=\biggl[1-\biggl(\frac{2}{1+T^2}\biggr)\biggr(\frac{Z_i}{m_i}\frac{\Delta t}{2}\biggr)^2B^2\biggr]u^{\theta-}(t)\nonumber\\
- \biggl(\frac{2}{1+T^2}\biggr)\biggr(\frac{Z_i}{m_i}\frac{\Delta t}{2}\biggr)\biggl(\frac{g}{J}\biggr)\text{g}_{\psi\psi}(t)u^{\psi-}(t),\\
u^{\zeta+}(t)=u^{\zeta-}(t)\quad\quad\quad\quad\quad\quad\quad\quad\quad\quad\quad\quad\quad\nonumber
\end{eqnarray}
where $T=(Z_iB/m_i)(\Delta t/2)$. We assume orthogonal basis vectors, so the off-diagonal terms of 
metric tensor are zero, and one can redefine the Jacobian as 
$J^2=\text{g}_{\psi\psi}\text{g}_{\theta\theta}\text{g}_{\zeta\zeta}$, and the diagonal components of the 
metric tensor as $\text{g}_{\psi\psi}=(\partial r/\partial\psi)^2$, $\text{g}_{\theta\theta}=r^2$,
$\text{g}_{\zeta\zeta}=(R_0+r\text{cos}\theta)^2$.
For $g=1$, and $B=(1+r/R_0\text{cos}\theta)^{-1}$, one can explicitly prove that
\begin{eqnarray}
{ u^{\psi+}}^2\text{g}_{\psi\psi}(t)+{ u^{\theta+}}^2\text{g}_{\theta\theta}(t)+{ u^{\zeta+}}^2\text{g}_{\zeta\zeta}(t)=\nonumber\\
{ u^{\psi-}}^2\text{g}_{\psi\psi}(t)+{ u^{\theta-}}^2\text{g}_{\theta\theta}(t)+{ u^{\zeta-}}^2\text{g}_{\zeta\zeta}(t)
\end{eqnarray}

i.e., the magnitude of velocity is unchanged during the rotation. In the third step, we
add the other half electric acceleration to the rotated vectors to obtain the velocity at time $(t+1/2)$

\begin{figure}
\centering
\includegraphics[width=0.5\textwidth ]{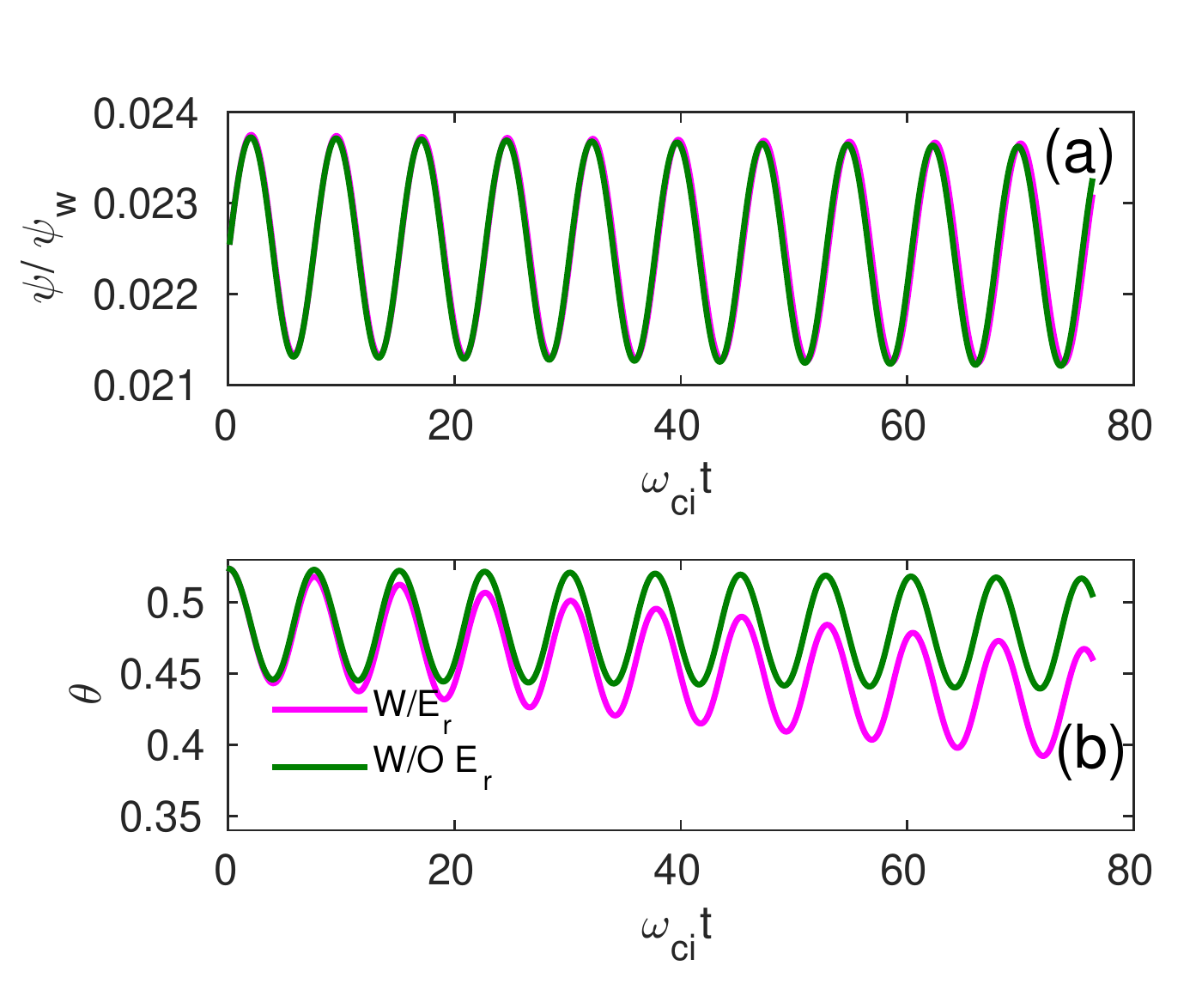}
\caption{\label{fig:epsart} Verification of $\vec{E}\times\vec{B}$  drift in toroidal geometry.
Time variations of (a) poloidal flux function and (b) poloidal angle
of the ion position. Here $\psi_w$ is the poloidal flux function
at the last closed flux surface.}
\end{figure}

\begin{figure*}
\centering
\includegraphics[width=0.9\textwidth ]{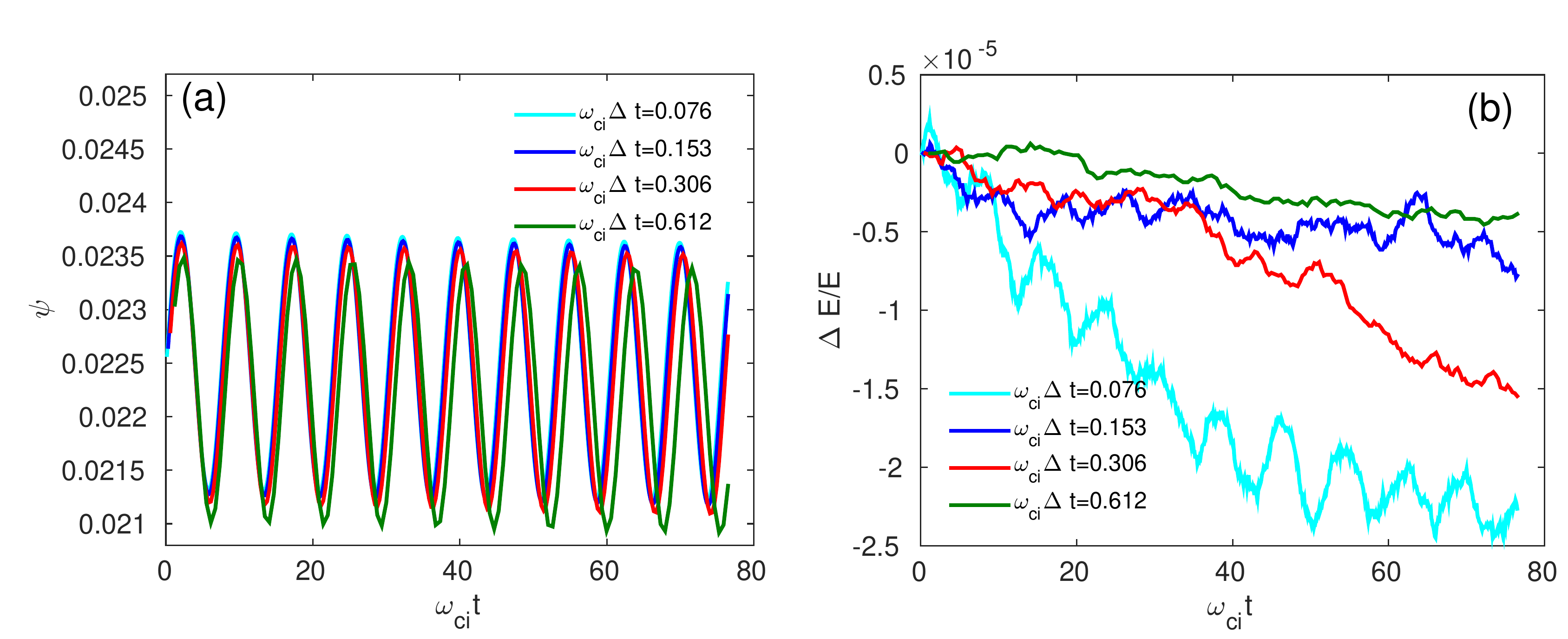}
\caption{\label{fig:epsart} Time step convergence of (a) poloidal flux function, and
(b) relative energy error of ion.}
\end{figure*}

\begin{equation}
 u^\alpha(t+1/2)=u^{\alpha+}(t)+\frac{Z_i}{m_i}\frac{\Delta t}{2}\vec{E}(t)\cdot\nabla\alpha(t)
\end{equation}
To update the particle position we need to recover $\vec{v}(t+1/2)$, which can be done 
through the following transformation (cf. Fig.2 dark purple arrow)
\begin{equation}
 v^\gamma(t+1/2)=\sum_{\alpha=\psi,\theta,\zeta}u^{\alpha}(t+1/2)\vec{e}_\alpha(t)\cdot\nabla\gamma(t+1/2)
\end{equation}
where $\gamma=\psi,\theta,\zeta$. However, the basis vector $\nabla\gamma(t+1/2)$ is still unknown, since
$\gamma(t+1/2)$ does not exist in standard leap-frog scheme. Here we use an estimator for $\gamma(t+1/2)$ as
\begin{equation}
 \gamma(t+1/2)=\gamma(t)+u^\gamma(t+1/2)\frac{\Delta t}{2}
\end{equation}
After we find the velocity at time $(t+1/2)$, we can update the particle position using the leap-frog scheme as
\begin{equation}
 \gamma(t+1)=\gamma(t)+v^\gamma(t+1/2)\Delta t
\end{equation}

In Eq.(27) we have the dot-product of two basis vectors at different time steps. We have evaluated this equation in the similar
way as described in Eqs. (20)-(22). 

To verify this cyclotron integrator we have carried out a single particle ion dynamics in toroidal geometry
with inverse aspect ratio $r/R_0=0.357$, and $\rho_i/r=0.0048$. First we calculate the time variations of the the 
poloidal flux function and the poloidal angle in the absence of electric field 
[cf. Fig.3 green line]. 
Secondly we introduce a radial electric field in the particle equation of motion. In the presence of
this static electric field, an ion will experience a $\vec{E}\times\vec{B}$ drift in the poloidal direction, 
as shown in Fig.3 (magenta line). These results between the theory and GTC simulations
are summarized in Table I. 

Fig. 4(a)-(b) show the time step convergence of the poloidal flux
function and the relative energy error of the marker particle. Fig. 4(a) shows
that poloidal flux function can converge with 40 time steps per cyclotron period
$(\omega_{ci}\Delta t=0.153)$. However, there is no such time dependent relation for the
calculation of the energy error.
Error in the energy arises mostly due to the decomposition of velocity during first and last 
steps of the Boris scheme and it is in the acceptable limit $(\sim 10^{-5})$.

In the above section we have discussed the time advancement of the
dynamical quantities like velocity and position of ion in the time centered manner. However, for 
the the self-consistent simulation we need to update particle
weight, electron guiding center and electric field.  We use the second order Runge-Kutta (RK) method, 
to advance these quantities, which can be described as follows. In our global simulation 
we use reflective boundary condition for the particle and fields.

\subsection{RK pusher for particle weight and electron dynamics}
Using the initial ion velocity at time $(t-1/2),$ the ion 
particle push module update the velocity up to (t+1/2). The 
velocity at time (t) is computed by linear average of velocities as follows
\begin{equation}
 v^\gamma(t)=\frac{v^\gamma(t-1/2)+v^\gamma(t+1/2)}{2}
\end{equation}
Now using the electric field and average velocity at time $(t)$, one can compute $(dw_i/dt)$
from Eq.(10). 
For the first step of the RK method we advance the particle weight from 
$w_i(t)$ to $w_i(t+1/2)$, and $w_e(t)$ to $w_e(t+1/2)$.
With the updated values of the source term (charge density), the field solver compute the electric field at 
time $(t+1/2)$ (cf. Fig.2 orange arrow). In the second step of RK we use the updated electric field
at $(t+1/2)$ for the advancement of the particle weight, and electron guiding center $(z_e)$ 
from $(t)$ to $(t+1)$ [cf. Fig.2 dark blue arrow].
Mathematically we can write these two steps as follows

 \renewcommand{\labelitemi}{$\blacksquare$}
 \renewcommand\labelitemii{$\square$}
 \begin{itemize}
   \item  First step (irk=1)
 \end{itemize}

\begin{eqnarray}
 \blacktriangleright\text{Ion pusher}\rightarrow w_i(t+1/2)=
 w_i(t)+\frac{dw_i}{dt}\biggr|_t\frac{\Delta t}{2},\quad\quad\quad\nonumber\\
 \left.
		\begin{aligned}
			\blacktriangleright\text{Electron}&\\
			\text{pusher} &
		\end{aligned}
	\right.
	\quad \Rightarrow
	\left\{
		\begin{aligned}
			&z_e(t+1/2)=z_e(t)+\frac{dz_e}{dt}\biggr|_t\frac{\Delta t}{2} \\
			&w_e(t+1/2)=w_e(t)+\frac{dw_e}{dt}\biggr|_t\frac{\Delta t}{2}\quad\quad\nonumber
		\end{aligned}
	\right.	\\
\bullet \quad\text{Solve field solver for} \quad\vec{E}(t+1/2)\quad\quad\quad\quad\quad\quad\quad\quad\quad\nonumber
\end{eqnarray}

\begin{itemize}
 \item Second step (irk=2)
\end{itemize}
\begin{eqnarray}
 \blacktriangleright\text{Ion pusher}\rightarrow w_i(t+1)=
 w_i(t)+\frac{dw_i}{dt}\biggr|_{(t+1/2)}\Delta t,\quad\quad\quad\nonumber\\
 \left.
		\begin{aligned}
			\blacktriangleright\text{Electron}&\\
			\text{pusher} &
		\end{aligned}
	\right.
	 \Rightarrow
	\left\{
		\begin{aligned}
			&z_e(t+1)=z_e(t)+\frac{dz_e}{dt}\biggr|_{(t+1/2)}\Delta t \\
			&w_e(t+1)=w_e(t)+\frac{dw_e}{dt}\biggr|_{(t+1/2)}\Delta t\quad\quad\nonumber
		\end{aligned}
	\right.	\\
\bullet \quad\text{Solve field solver for} \quad\vec{E}(t+1)\quad\quad\quad\quad\quad\quad\quad\quad\quad\quad\nonumber
\end{eqnarray}

\begin{table}
\caption{\label{tab:table1}Comparison of ion cyclotron motion including $\vec{E}\times\vec{B}$
drift from simulation and theory}
\begin{ruledtabular}
\begin{tabular}{lcr}
Parameter & Theory &GTC Simulation\\
\hline
\\
Gyro radius &8.415$\times10^{-3}$(m) & 8.46$\times10^{-3}$(m)  \\
Gyro frequency &4.557$\times10^6$(rad/sec) & 4.538$\times10^{6}$(rad/sec)  \\
$\vec{E}\times\vec{B}$ drift &1.815$\times10^4$(m/s) &  1.785$\times10^4$(m/s) \\
\end{tabular}
\end{ruledtabular}
\end{table}

\begin{figure*}
\centering
\includegraphics[width=0.9\textwidth]{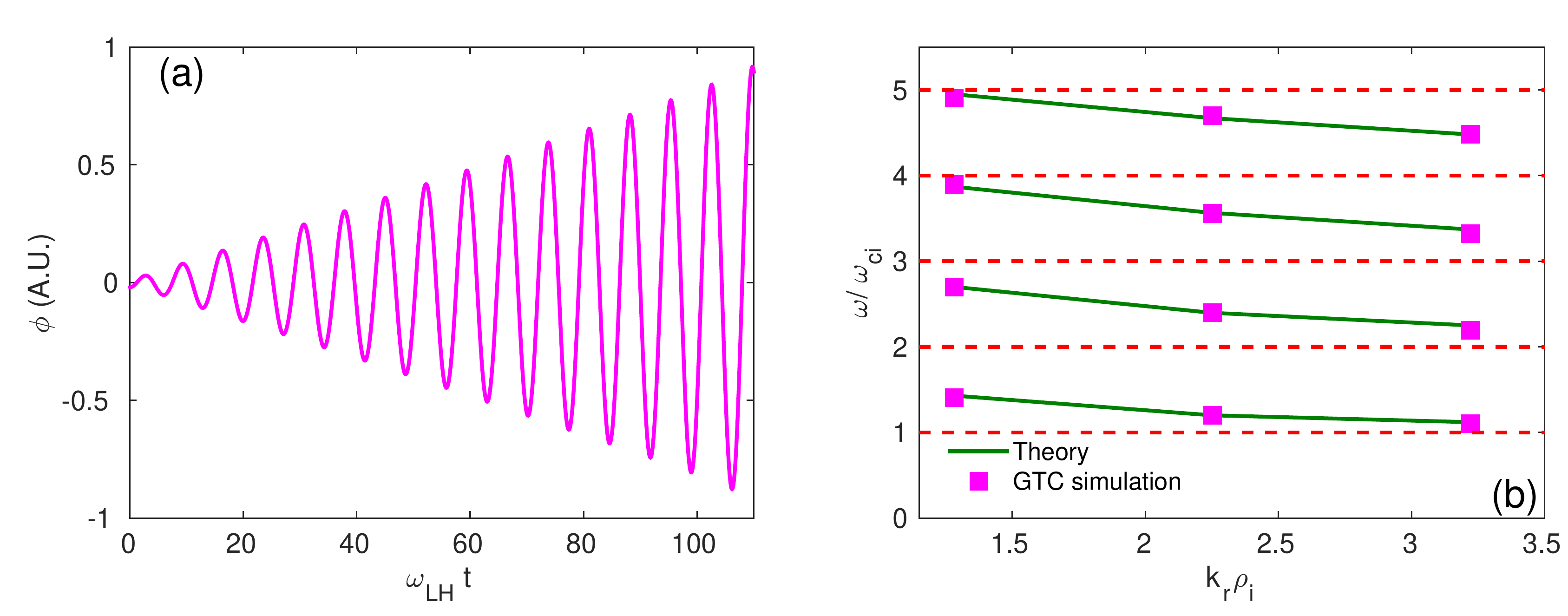}
\caption{\label{fig:epsart} Verification of Normal modes in toroidal geometry.
(a) Time history of LH wave amplitude excited by antenna, 
(b) comparison of IBW dispersion relation between the
analytical solution and the GTC simulations for the first four harmonics.}
\end{figure*}
\begin{figure}
\centering
\includegraphics[width=0.5\textwidth]{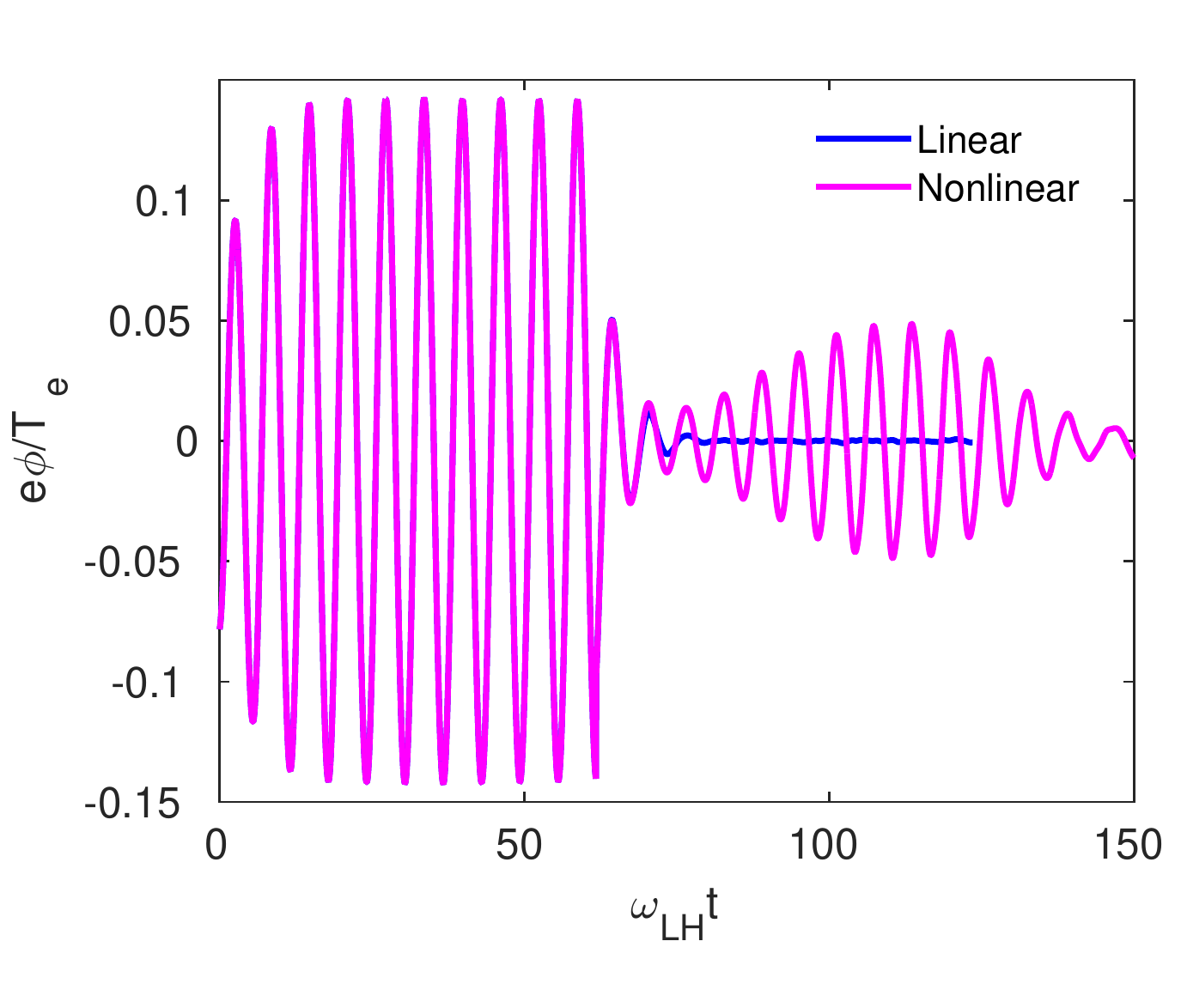}
\caption{\label{fig:epsart} Nonlinear GTC simulation of LH wave exhibits oscillation in wave amplitude
(magenta line), while linear simulation shows exponential decay (blue line).}
\end{figure}

\section{Linear Verification of Normal modes and nonlinear particle trapping}
In this section, we will discuss the electrostatic normal modes with $k_\parallel=0$ as 
a  benchmark of toroidal Boris scheme in the linear simulation.
The general dispersion relation of the normal mode in uniform plasma 
can be written as \cite{Liu86}
\begin{equation}
1+\chi_j=0
\end{equation}
For a  Maxwellian background, one can write down the susceptibility as

\begin{equation}
\chi_j=-\sum_{j=e,i} \frac{1}{k^2\lambda_{Dj}^2} \sum_{l=1}^\infty \frac{2l^2\omega_{cj}^2}{\omega^2-l^2\omega_{cj}^2}
I_l(b_j)e^{-b_j}
\end{equation}
where $b_j=k_\perp^2\rho_j^2$, $\rho_j=\sqrt{(T_j/m_j)}/\omega_{cj}$, $\lambda_{Dj}^2=\epsilon_0T_j/n_{0j}e^2$,
and $\omega_{ci}=Z_iB/m_i$.
For normal modes (LHW, IBW), we have $|\omega/\omega_{ce}|\sim \mathcal{O}(\omega/\omega_{ce})\ll 1$ and
$k_\perp \rho_e\ll 1$, hence the electron susceptibility is dominated by $l=1$ term.
So the above Eq (32) becomes
\begin{equation}
 1+\frac{\omega_{pe}^2}{\omega_{ce}^2}-
 \frac{1}{k_{\perp}^2\lambda_{Di}^2}\sum_{l=1}^{\infty}\frac{2l^2\omega_{ci}^2}
 {\omega^2-l^2\omega_{ci}^2} I_{l}(b_i)e^{-bi} =0
\end{equation}
For long wavelength limit $k\rightarrow 0^+$, with $\omega_{ci}\ll\omega\ll\omega_{ce}$, the frequency of the LH is 
\begin{equation}
\omega_{LH}^2=\omega_{pi}^2\biggl(1+\frac{\omega_{p}^2}{\omega_c^2}\biggr)^{-1}
\end{equation}

We use an artificial antenna to excite these modes and to verify the mode
structure and frequency in our simulation. The electrostatic potential of the 
antenna can be written as:
\begin{equation}
\phi_{ext}=\phi_0\text{sin}(k_r r)\text{cos}(\omega_0t)\text{cos}(m_0\theta-n_0\zeta)
\end{equation}


In our simulation, the inverse aspect ratio of the tokamak is $r/R_0= 0.018$, $\rho_i/r=0.002$, and
the background plasma density is uniform with a uniform temperature. 
We use poloidal and toroidal mode filters to select m=0, n=0 modes. 
For the lower hybrid simulation, $\omega_{pi}=145.2\omega_{ci}$,,
$\omega_{pe}=6242.8\omega_{ci}$, $\omega_{ci}\Delta t\simeq1.33\times 10^{-3}$,
$m_e/m_i=5.44618\times 10^{-4}$,
and particles per wavelength are 
$8\times 10^6$. 
We carry out the scan with different antenna frequencies, 
and find out the frequency in which the mode response has the maximum growth of 
the amplitude. That frequency is then identified as the eigenmode frequency of the system.
Fig. 5(a) shows the time history of the 
lower hybrid wave amplitude for the antenna frequency $\omega_0=41.0\omega_{ci}$, which
agrees well with the analytical frequency $40.9\omega_{ci}$. Also the 
amplitude of the LH wave increases linearly with time, since there is no damping due
to $k_\parallel=0$. To avoid the boundary effects we consider $(\Delta/\rho_i=41.13)$, 
where $\Delta$ is the width of the simulation domain.

\begin{figure}
\centering
\includegraphics[width=0.5\textwidth]{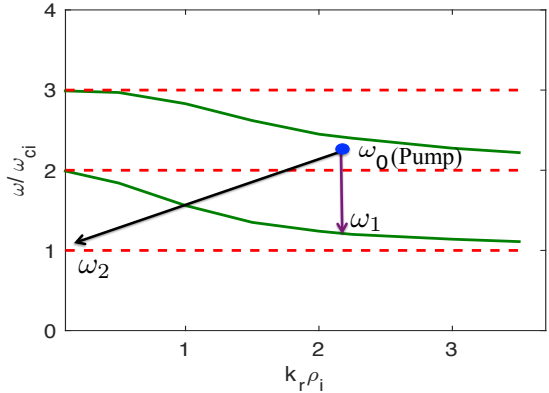}
\caption{\label{fig:epsart} Schematic of an IBW parametric decay process. Pump wave $(\omega_0,\vec{k}_0)$ 
decays into an IBW side band  $(\omega_1,\vec{k}_1)$ and
an ion cyclotron quasi mode (ICQM) $(\omega_2,\vec{k}_2)$. 
Green lines represent the theoretical dispersion curve of the IBW
for first two harmonics. }
\end{figure}
\begin{figure*}
\includegraphics[width=0.9\textwidth]{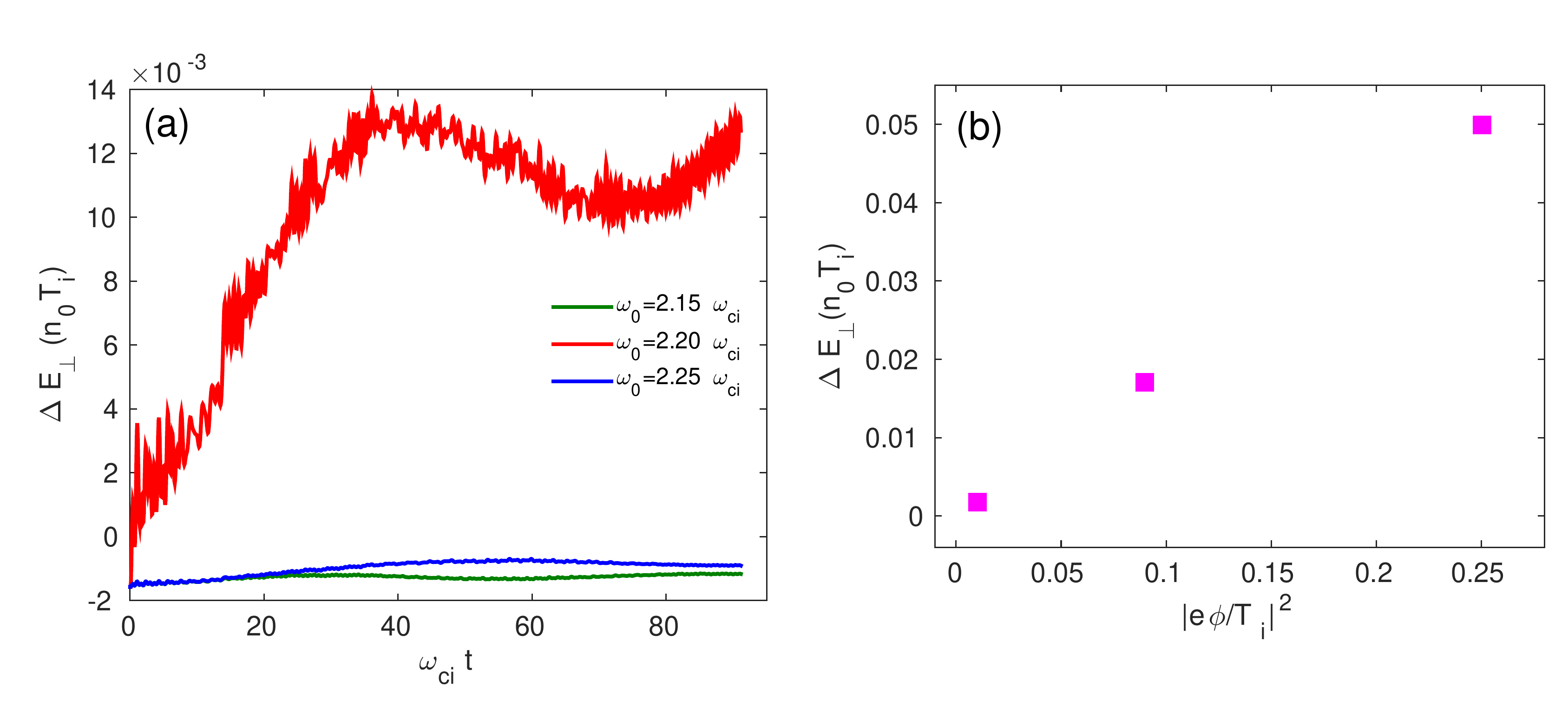}
\caption{(a) Time history of change of perpendicular kinetic energy of ion 
for different pump wave frequency, and (b) change in kinetic energy of ion as a function of
intensity of the pump wave.}
\end{figure*}


 Similarly we carried out the simulation of the IBW waves for the first four
harmonics $(l=1-4)$. In this simulation by changing the plasma density we consider $\omega_{pi}=10.01\omega_{ci}$,
$\omega_{pe}=422.3\omega_{ci}$, $\Delta/\rho_i=19.45$, $\omega_{ci}\Delta t\simeq0.055$,
and particles per wavelength are 
$8\times 10^6$. Fig. 5(b) shows a good agreement between the analytical and
GTC simulation results of the IBW frequency. 

\begin{figure}
\includegraphics[width=0.5\textwidth]{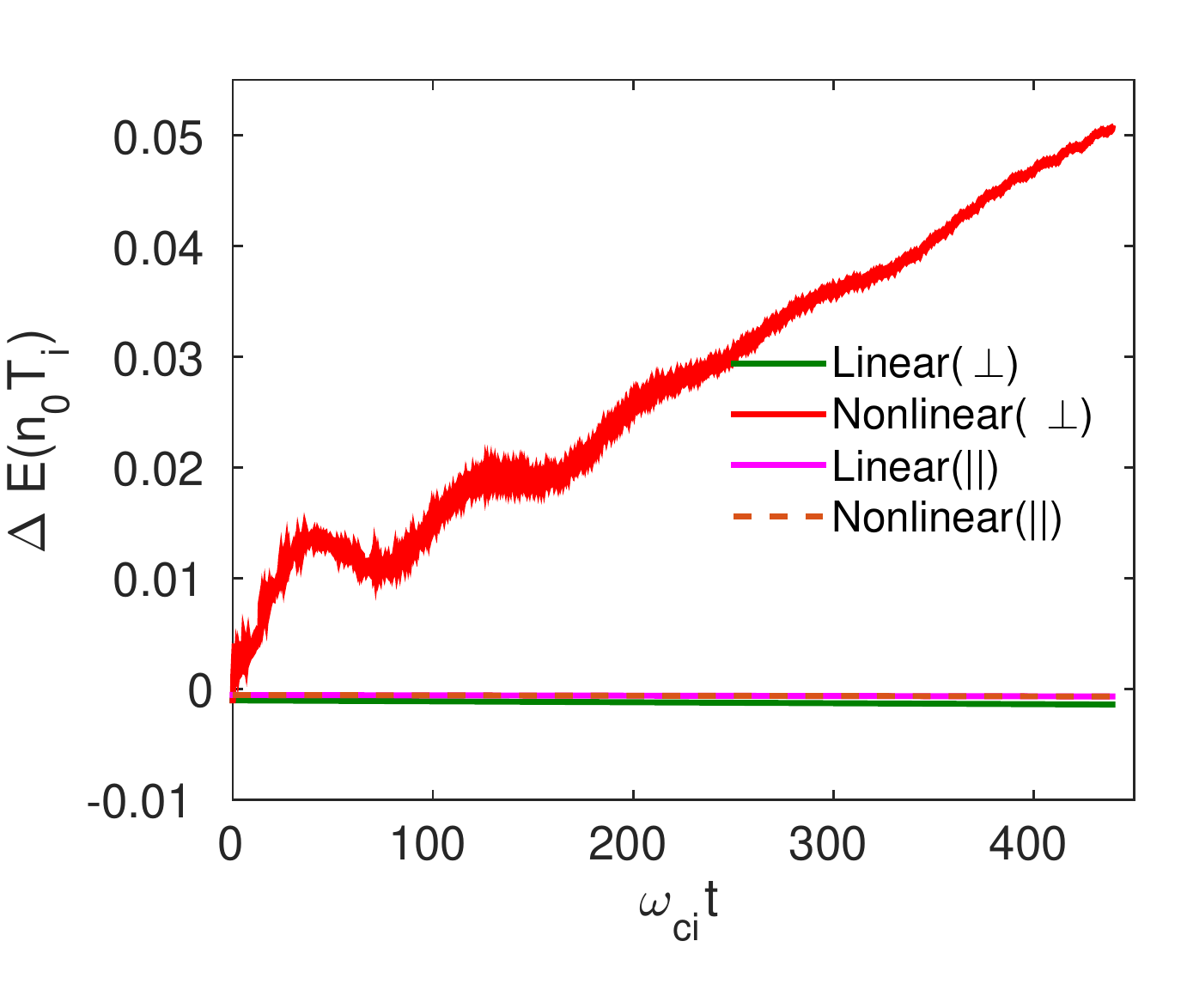}
\caption{Time history of change of kinetic energy of ion. Green and magenta lines
represent the energy change during linear simulation in the perpendicular and 
parallel direction, respectively. Red and dotted lines indicate the energy change
during nonlinear simulation in the perpendicular and 
parallel direction, respectively.}
\end{figure}


As a first step in developing this nonlinear toroidal particle simulation model, we carry out the 
nonlinear GTC simulation of electron trapping by the LH wave with a large 
amplitude [cf. Fig. 6] in cylindrical geometry.
Initially the linear lower hybrid eigen mode $(m=4$, and $n=1)$ is excited using an artificial antenna.
After the wave amplitude reaches the plateau regime, we turn off the antenna, and the
wave decays exponentially due to the Landau damping on electrons in the linear
simulation (blue line).\cite{0741-3335-56-9-095020} The linear damping rates obtained from the theory
$(0.3\omega_{LH})$ agree well with the simulation $(0.31\omega_{LH})$.
However, in the nonlinear simulation, the resonant electrons can be trapped by the electric field of 
the wave. The wave amplitude become oscillatory with a frequency equal to the trapped electron oscillation 
frequency (magenta line). The bounce frequencies $\omega_b=k_\parallel v_{the}\sqrt{e\phi/T_e}$ 
are close to the analytical values (Table II). During the antenna excitation of LH wave eignemode,
the particle dynamics are linear for both linear and nonlinear simulation.

\begin{table}
\caption{\label{tab:table1}Comparison of bounce frequency of nonlinearly
trapped particle in LH wave simulation}
\begin{ruledtabular}
\begin{tabular}{lcr}
$e\phi/T_e$ & Theory &GTC Simulation\\
\hline\\
0.00885 &2.91$\omega_{ci}$ & 3.04$\omega_{ci}$  \\
0.04866 &6.8$\omega_{ci}$ & 6.7$\omega_{ci}$  \\

\end{tabular}
\end{ruledtabular}
\end{table}


\section{PDI of ion cyclotron wave and Nonlinear Ion heating}
Magnetized plasma supports a large number of electrostatic and electromagnetic 
modes. 
When wave energy is still relatively low, these modes are mutually independent and 
represent a description for the response of the plasma to local perturbation and 
external field. However, at higher amplitudes these modes are coupled and 
exchange momentum and energy with each other through the coherent wave 
phenomenon , e.g., parametric decay instabilities (PDI). In this process the 
pump wave $(\omega_0,\vec{k}_0)$ decays into two daughter waves or one 
daughter wave $(\omega_1,\vec{k}_1)$ and a quasimode $(\omega_2,\vec{k}_2)$ pair. 
The selection rule for this decay process is given by

\begin{equation}
\omega_0=\omega_1+\omega_2 \quad\quad \quad \vec{k}_0=\vec{k}_1+\vec{k}_2
\end{equation}

Looking at Figure 7, we see that in this case, the pump wave can decay 
into one daughter with a near zero wavenumber and another wave with nearly the 
same wavenumber as the pump wave. This process is considered to be
the most probable non-resonant decay channel of PDI in the ion cyclotron heating regime.
Possible decay channels in the ion cyclotron range
of frequency in experiments have been discussed by Porkolob.\cite{Porkolab199093}
In Figure 7, we show the IBW dispersion 
relation (green lines) for the same parameter 
as in Figure 5(b). In our simulation, the pump wave itself is not an IBW,
and the value of $\omega/\omega_{ci}$ at the antenna position is 2.25, as indicated in Fig.7.
When $\omega_0=\omega_{\text{IBW}}+\omega_{ci}$ i.e., the frequency shift of the wave is close
to the ion cyclotron frequency, the wave is strongly damped on ions and known as
ion cyclotron quasi-mode (ICQM). 

In experiment, a pump wave of fixed frequency passing through the nonuniform
plasma  density and temperature experiences 
the variation of the wave vector to satisfy the dispersion relation. 
As a result of this inhomogeneity, layers may exist where selection rules
of mode-mode coupling are easily satisfied.  However, in our simulation 
plasma density and temperature are uniform and the wave vectors of the pump wave and
the sideband wave are chosen by the antenna. To satisfy the frequency matching 
condition in our simulation, we scan the pump wave frequencies with fixed wavevector. 
The energy transfer from the wave to ion (nonlinear ion heating) is maximum,
when the frequency selection of parametric decay, $\omega_1=\omega_{\text{IBW}}$ and
$\omega_2=\omega_{ci}$ are
satisfied (cf. Fig.8(a) red line). Otherwise the energy transfer is negligible 
(cf. Fig.8(a) green and blue lines). 
Our simulations are all electrostatic and we consider only m=0, n=0 mode.
We consider $\omega_{pi}=10.01\omega_{ci}$,
$\omega_{pe}=422.3\omega_{ci}$, $\Delta/\rho_i=19.45$, $100$ radial grid points per
wavelength, 200 particles per cell. The simulation time step $(\omega_{ci}\Delta t\simeq0.02)$ is
sufficient to resolve 
IBW, ion cyclotron wave and pump wave dynamics. In this case the particles trajectory 
are described by the perturbed electric field in addition to the equilibrium 
magnetic field [cf. IIA]. Fig.8(b) shows that the temperature of the hot ions
increases linearly as the amount of rf power increases, since the kinetic energy
of the ion [$(1/2)m_i(U/c_s)^2$] is proportial to $(e\phi/T_i)^2$, where 
$U/c_s=k_{0\perp}\rho_i(e\phi/T_i)$ is the normalized ion velocity, and $c_s=\sqrt{T_e/m_i}$ is
the ion sound speed. We measure the energy of the ion after 400 ion cyclotron periods. 
This simulation time is long enough to excite the daughter waves for the prominent
PDI phenomenon. 

Fig.9 shows that ion heating 
takes place only in the perpendicular direction. The ion temperature in the 
parallel direction does not change. Since the wave heating affects
predominantly the perpendicular ion distribution, which is consistent with the results
observed in the scrape-off 
layer (SOL) of DIII-D,\cite{0029-5515-33-5-I08} 
Alcator-C Mod,\cite{:/content/aip/journal/pop/9/4/10.1063/1.1456531} 
and HT-7\cite{0741-3335-43-9-305} experiments during the IBW heating and 
high harmonic fast waves heating in 
NSTX.\cite{:/content/aip/journal/pop/12/5/10.1063/1.1871953}  
This parasitic absorption of the wave energy degrade the performance
of ion Bernstein and ion cyclotron harmonics resonance heating.
However, our simulations are limited to the core region only.
In our nonlinear simulation the ponderomotive effect is absent, since we consider the plasma response in
the $r$ direction only.
With the present simulation setup, the amount of power transferred 
to the sideband (IBW) has not been measured directly, but the energy of the 
quasi-mode is measured from the particle diagnosis. During the linear simulation,
wave particle interaction can be possible only through the linear damping, which is negligible
compared to the nonresonant damping.

\section{Discussion}
In summary, nonlinear global toroidal particle simulations have been developed using 
fully kinetic ion and drift kinetic electron to study the electron trapping by LHW and 
parametric decay process of ion cyclotron range of frequency (ICRF) waves in
uniform core plasma. We verify our simulation results with the linear dispersion
relation. In the nonlinear simulation of LH wave, we find that the amplitude of the
electrostatic potential oscillates with a bounce
frequency, which is due to the
wave trapping of resonant electrons. We also find the nonlinear anisotropic ion heating
due to nonresonant three wave coupling.
One must mention here that in tokamak scenario with non-uniform density and
temperature, energy density of the wave can begin to approach the thermal energy
in the edge.  Since, in the edge region the densities and temperatures are factors
of $10^2$ to $10^3$ lower than in the core produce strong ponderomotive effects and
parametric decay physics.

\begin{acknowledgments}
Author AK would like to thank Dr. R. B. White for his useful suggestions. 
This work is supported by PPPL subcontract number S013849-F, US Department of Energy (DOE)
SciDAC GSEP Program and China National Magnetic Confinement Fusion Energy Research Program,
Grant No. 2013GB111000, and 2015GB110003. Simulations
were performed using the super computer resources of the Oak Ridge Leadership
Computing Facility at Oak Ridge National Laboratory (DOE Contract No. DE-AC05-00OR22725),
and the  National Energy Research Scientific Computing 
Center (DOE Contract No. DE-AC02-05CH11231).
\end{acknowledgments}

\nocite{*}
\bibliography{Kuley-15}

\end{document}